\documentstyle[preprint,aps]{revtex}
\input epsf.tex
\def\DESepsf(#1 width #2){\epsfxsize=#2 \epsfbox{#1}}
\begin{document}
\preprint{\vbox{\hbox{}}}
\draft
\title{
Effects of Extra Dimensions on Unitarity and Higgs Boson Mass}
\author{Xiao-Gang He\footnote{E-mail: hexg@phys.ntu.edu.tw}}
\address{
Department of Physics, National Taiwan University, Taipei, Taiwan 10617, R.O.C.
}
\date{May 1999}
\maketitle
\begin{abstract}
We study the unitarity constraint on the two body Higgs boson elastic
scattering in the presence of extra dimensions. 
The contributions
from exchange of spin-2 and spin-0 Kaluza-Klein states 
can have large effect on the partial wave amplitude. Unitarity 
condition restrict the maximal allowed value for the 
ratio $r$ of the 
center of mass energy to the gravity scale to be less than one. 
Although the constraint on the standard Higgs boson mass for $r$ of order 
one is considerably relaxed, for small $r$ the constraint 
is similar to that in the Standard Model. The resulting bound on the Higgs 
boson mass is not dramatically altered if 
perturbative calculations are required to be valid up to the maximal allowed 
value for $r$.
\end{abstract}
\newpage
It has recently been proposed
that gravitational effects can become large at a scale
$M_S$ near the weak scale due to effects from extra dimensions\cite{1,2},
quite different from the traditional thought that gravitational effects
only become large at the Planck scale $M_{Pl} \sim 10^{19}$ GeV.
In this proposal the total space-time is $D=4 + n$.
The relation between the scale $M_S$ and the scale
$M_{Pl}$, assuming all extra dimensions are compactified with the
same size R, is given by $M^2_{Pl} \sim R^n M^{2+n}_S$. With
$M_S$ near a TeV and n= 1, $R$ would be too large. 
However, with n larger than or equal to
2, the theory is not ruled out for $M_S \sim 1$ TeV. 
The lower bound for $M_S$ is
constrained, typically, to be of order one TeV from present
experimental data\cite{3,4,5,6,7}. 
Future experiments will provide more stringent
constraints\cite{3,4,5,6,7}. 
There are many interesting phenomena due to the presence of extra 
dimensions\cite{3,4,5,6,7,8,9}. 
In this paper we study effects from extra dimensions on the unitarity condition
of partial wave amplitude in elastic two body Higgs boson scattering, and to study 
implications for the validity of perturbative 
calculations and for 
allowed Higgs boson mass.

In the minimal Standard Model (SM) there is a neutral Higgs boson $H$
resulting from spontaneous symmetry breaking of $SU(2)_L\times
U(1)_Y$ to $U(1)_{em}$ due to the Higgs  mechanism. The mechanism
for spontaneous symmetry breaking is not well understood. There is
no experimental evidence favoring any particular mechanism, such
as the Higgs mechanism. The discovery of the Higgs boson and
understanding of its properties are fundamentally
important\cite{10}. 
Many
methods have been proposed to produce and to study the properties
of Higgs bosons\cite{10}. 
One of the most important issue is its mass.
At present the lower bound on SM Higgs boson mass
$m_H$ is set by LEP II to be 95.5 GeV at 95\% C.L.\cite{11}. 
There are many theoretical constraints on the Higgs boson masses.
The constraint from unitarity of partial wave amplitudes of longitudinal 
gauge boson and/or Higgs boson scatterings provide some of the interesting
upper bounds on the mass\cite{12,13,14,15}. 

In the presence of extra dimensions there are additional
contributions to gauge and/or Higgs boson scatterings 
due to exchanges of spin-2 and spin-0 graviton excitations (the KK states). 
The effects of these KK states can affect the partial wave amplitudes
significantly if the ratio $r$ of the center of mass frame energy 
$\sqrt{s}$ and the gravity scale $M_S$ is close to or larger than one.
Unitarity condition for partial wave amplitude constrain $r$ to be less than
one if perturbative calculations are valid.
The 
allowed range for Higgs boson mass can also be different from that in the SM 
depending on the value of $r$. 
We find that effects from extra dimensions affect the Higgs boson 
scattering $HH\to HH$ the largest. In the following we will concentrate on 
this process, and
will comment on other processes at the end.

In the SM, the scattering amplitude for the process $HH\to HH$ at tree level
is given by\cite{13}

\begin{eqnarray}
M_{SM}(s,t) &=& 3\sqrt{2} G_F m_H^2 ( 1 + {3m_H^2\over s-m_H^2}
+ {3m_H^2\over t-m_H^2} + {3m_H^2\over u-m_H^2}),
\end{eqnarray}
where $s$, $t$ and $u=4 m_H^2 -s-t$ are the Mandelstam variables.

From the above expression, one obtains the J=0 partial wave amplitude\cite{13} 

\begin{eqnarray}
a_0^{SM} &=& {1\over 16\pi}\left ( {4 p_i p_f\over s}\right )^{1/2}
 {1\over s- 4 m_H^2}
\int^0_{-(s-4m_H^2)} M_{SM}(s,t) dt\nonumber\\
&=& {G_F m_H^2\over 8\sqrt{2} \pi} \sqrt{1- {4m_H^2\over s}}
[ 3 + {9m_H^2\over s- m_H^2}
- {18 m_H^2\over s-4m_H^2} \ln ({s\over m_H^2} -3)].
\end{eqnarray}
In the above $p_{i,f}$ are the momentum of the initial and final Higgs 
boson in the center of mass frame, respectively.

The Higgs boson mass is constrained if one requires the absolute value of
$a_0$ not to violate the unitarity condition. There are many discussions of
how to implement unitarity conditions\cite{14}. 
For our purpose of demonstrating possible large effects of extra dimensions, 
we use a weak condition $|a_0| <1$ and work with tree level amplitude 
to obtain conservative bound.
Applying this condition
for $s\gg m_M^2$, one obtains

\begin{eqnarray}
m_H^2 < {8\sqrt{2} \pi \over 3 G_F} = 1010 \mbox{GeV}.
\end{eqnarray} 
If $m_H$ is substantially less than the above bound, the magnitude of the 
amplitude is well within the bound everywhere.

With extra dimensions, there are new contributions to $HH\to HH$ due to exchange of KK states. Using the 
Feymann rules derived in Ref\cite{6}, we obtain

\begin{eqnarray}
M_{NEW}(s,t)&=&
{\kappa^2\over 2}
\left 
\{ {1\over s - m^2_l} [ (2m_H^2-t)^2 + (2m_H^2-u)^2 -{2\over 3} (s-2m_H^2)^2 
- {4\over 3} m_H^2 s]\right .\nonumber\\
&+& {1\over t - m^2_l} [ (2m_H^2-s)^2 + (2m_H^2-u)^2 -{2\over 3} (t-2m_H^2)^2 
- {4\over 3} m_H^2 t]\nonumber\\
&+&\left . {1\over u - m^2_l} [ (2m_H^2-s)^2 + (2m_H^2-t)^2 -{2\over 3} (u-2m_H^2)^2 
- {4\over 3} m_H^2 u]\right \}\nonumber\\
&+&\kappa^2\left \{ {2(n-1)\over 3(n+2)}
[ {(s+2m_H^2)^2\over s-m_H^2} + {(t+2m_H^2)^2\over t-m_H^2}
+ {(u+2m_H^2)^2\over u-m_H^2}]\right \}.
\end{eqnarray}
The first and the second terms are due to exchanges of spin-2 and spin-0 KK 
states, respectively.

Summing over all intermediate KK states and projecting out the 
J=0 partial wave amplitude, we obtain

\begin{eqnarray}
a_0^{NEW} &=& \sqrt{1-{4m_H^2\over s}} \left \{ 
{2\over 3(n+2)} + {11s-12 m_H^2\over 3nM_S^2}\right .\nonumber\\
&-&\left . {2\over 3(s-4m_H^2)} [ M_S^2 Fn(4) + (6s-8m_H^2) Fn(2) + 
{6s(s-4m_H^2)+16 m_H^2 \over M_S^2}Fn(0)]\right \}
\nonumber\\ 
&+&{4(n-1)\over 3(n+2)} \left 
\{{1\over 2} (s+2m_H^2)^2 {s^{n/2-1}\over M_S^{n+2}} (-i\pi + 2 In(M_S/\sqrt{s}))
+ {2\over n+2} + {12m_H^2-s \over n M_S^2}\right .\nonumber\\
&-&\left . {2\over s-4m_H^2} [M_S^2 Fn(4) + 4 m_H^2 Fn(2) + 
{4 m_H^4\over M_S^2} Fn(0)]\right \}\sqrt{1-{4m_H^2\over s}},
\end{eqnarray}
where

\begin{eqnarray}
In(x) = \int^x_0 {y^{n-1}\over 1-y^2}dy,\;\;
Fn(\delta) = \int^1_0 x^{n+\delta -1} \ln[({s-4m_H^2\over M_S^2} + x^2)/x^2] dx.
\end{eqnarray}
In the above we have used $\kappa^2 = 16\pi G_N =16\pi
 (4\pi)^{n/2} \Gamma(n/2) R^{-n} M_S^{-(n+2)}$ 
as the convention for $M_S$.

In the expression for $a_0^{NEW}$ there are several constant terms which look dangerously large 
are all canceled by terms proportional to $Fn(4)$. In the large $M_S$ limit, $a_0^{NEW}$ is 
proportional to $1/M_S^4$ and 
approaches zero as $M_S$ goes to infinity with $s$ and $m_H$ kept finite.
In this limit the theory reduces to the
SM. However, when the 
ratio $\sqrt{s}/M_S$ approaches one, the real and imaginary parts of 
$a^{NEW}_0$ both 
become of order one and can violate the
unitarity condition $|a_0|<1$ even if $m_H$ is small. 
This indicates that the applicability of the effective 
theory, perturbatively, should be in the range $\sqrt{s} < M_S$.
We remark that $a_0^{NEW}$ only becomes
sensitive to $m_H$ for small $M_S$. With sufficiently large $M_S$, $a_0^{NEW}$ 
by itself does not provide 
interesting information
for $m_H$. However since the SM is sensitive to Higgs boson mass, 
the total contribution will provide information about $m_H$. 
To have a better understanding of the details we consider $|a_0|$ 
as a function of $r=\sqrt{s}/M_S$ and $m_H$ for 
four typical cases: a) $M_S = 5$ TeV and n=2; b) $M_S = 5$ TeV and n=7; 
c) $M_S= 100$ TeV and n=2; and d) $M_S = 100$ TeV and n= 7, for 
illustrations. 
For cases c) and d), the results reduce to the SM if $\sqrt{s}$ is not too close to
$M_S$, while for cases a) and b) the effects from extra dimensions can be large.
The results are shown in Figs. 1 and 2.

From Figs. 1 and 2, we see that in all cases when $r$ approaches one, $a_0$ becomes large and
can violate the unitarity condition setting the purterbative range (the maximal allowed 
range $r_{max}$ for $r$ where $|a_0| = 1$) of the theory to
be: 0.81, 0.96, 0.81 and 0.96
 for cases a), b), c) and d) with $m_H=100$ GeV, respectively.
With larger $m_H$, the allowed range for $r_{max}$ can be larger, as 
can be seen from Figs. 1 and 2, 
due to cancellation 
between contributions from the SM and extra dimensions, and the total $a_0$ 
is sensitive to $m_H$ as mentioned before.
As long as $\sqrt{s}$ is much 
larger than $2 m_H$, $r_{max}$ is not sensitive to $M_S$ but sensitive to
the number of extra dimensions n. 
We have studied different value for n up to 7 and find that in all cases, 
$r_{max}$ is constrained to be less than one, but varies from case to case. 
Violation of unitarity condition for $\sqrt{s} >M_S$ is not a big surprise
because $M_S$ serves as a cut-off where gravity becomes strong. 
The calculation here provides a 
specific example. We stress, however, that results obtained using perturbative
calculations with $\sqrt{s}$ close to $M_S$ are not reliable.

It has been shown\cite{14} that
in the SM at one loop level, for a given 
Higgs boson mass there is a critical scale $\sqrt{s_c}$ 
when $s $ is larger than $s_c$ the unitarity condition is violated 
($s_c$ decreases as $m_H$ increases).
With extra dimensions, the energy scale $s_{max}$ 
below which unitarity condition
is reached earlier for small $M_S$ and $m_H$.
If stronger requirement on $|a_0|$ is made as discussed in Ref.\cite{14}, 
the allowed $r_{max}$ will be smaller. The bound we obtained is a conservative
one.
 
The amplitude $a_0$ is also sensitive to the Higgs boson mass $m_H$ as mentioned earlier. 
In all cases the allowed range for the Higgs boson masses are altered,
in some cases the cahnge can be dramatical, such as case c), 
in certain ranges of $r$. 
For small $r$ the bound on the Higgs boson mass is similar to that in the 
SM. 
Higgs boson mass bound with smaller $r$ is tighter than
other ranges except $r$ close to its maximal as can be seen from Figs. 1 and 2.
The bound on Higgs boson mass must be the smallest in the whole 
valid range of $r$ for perturbative calculation. Therefore the 
bound on the Higgs boson mass is not dramatically affect compared with
SM case which is effectively determined with small value for $r$ in our 
cases.

We also checked some other gauge and/or Higgs boson scattering processes. 
For example,
the contribution  to $HH \to VV$ from extra dimensions 
in the limit neglecting the mass of the gauge boson V, is given by

\begin{eqnarray}
a^{NEW}_0 (HH \to VV)
= {1\over 12} {s^{n/2-1}\over M_S^{n+2}} [-i\pi + 2 In(M_S/\sqrt{s})]
s^2 \left ( 1- {4m_H^2\over s}\right )^{5/4}.
\end{eqnarray}
This is smaller than that for $HH\to HH$. 
Unitarity consideration for 
$HH\to HH$ obtain stronger constraint.

In conclusion we have shown that effects from extra dimensions can have 
large contributions to 
$HH\to HH$. 
The range valid
for perturbative calculations with extra dimensions is limited by the 
scale $M_S$. Results obtained with $\sqrt{s} > M_S$ are not reliable.
Although Higgs boson mass bound can be drastically affected for 
certain ranges of $\sqrt{s}$, the overall bound is not modified 
significantly if one requires perturbative calculation for $HH\to HH$ to 
be valid up to maximal allowed value of $r$. 

This work is supported in part by the National Science Council of R.O.C under
Grant NSC 88-2112-M-002-041.

\begin{figure}[htb]
\centerline{ \DESepsf(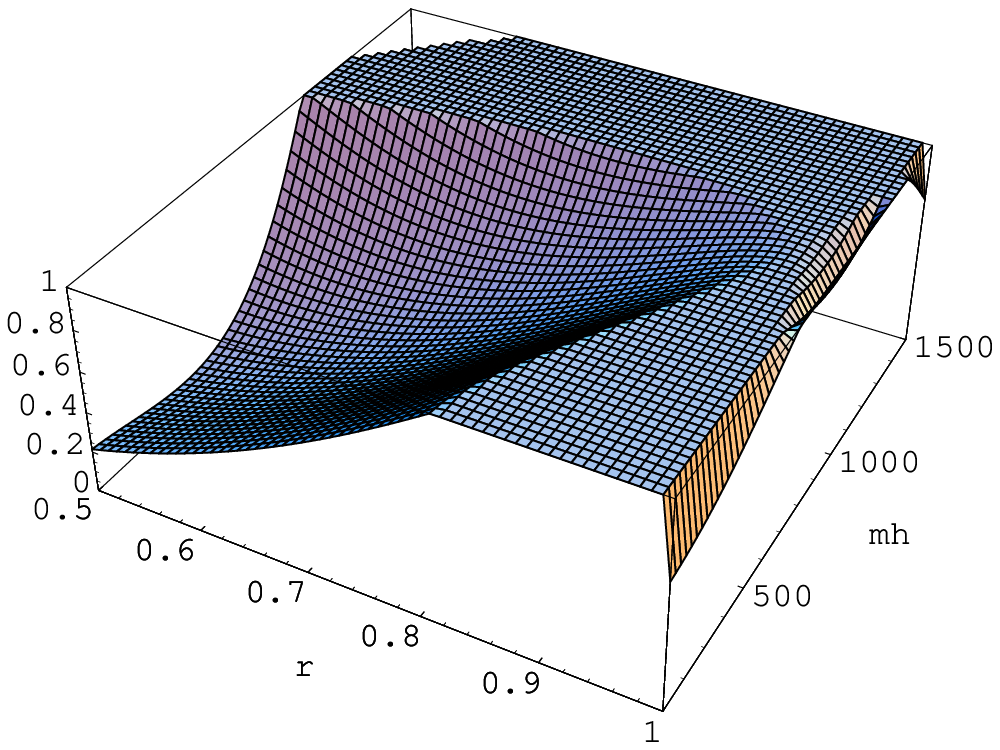 width 7 cm) \DESepsf(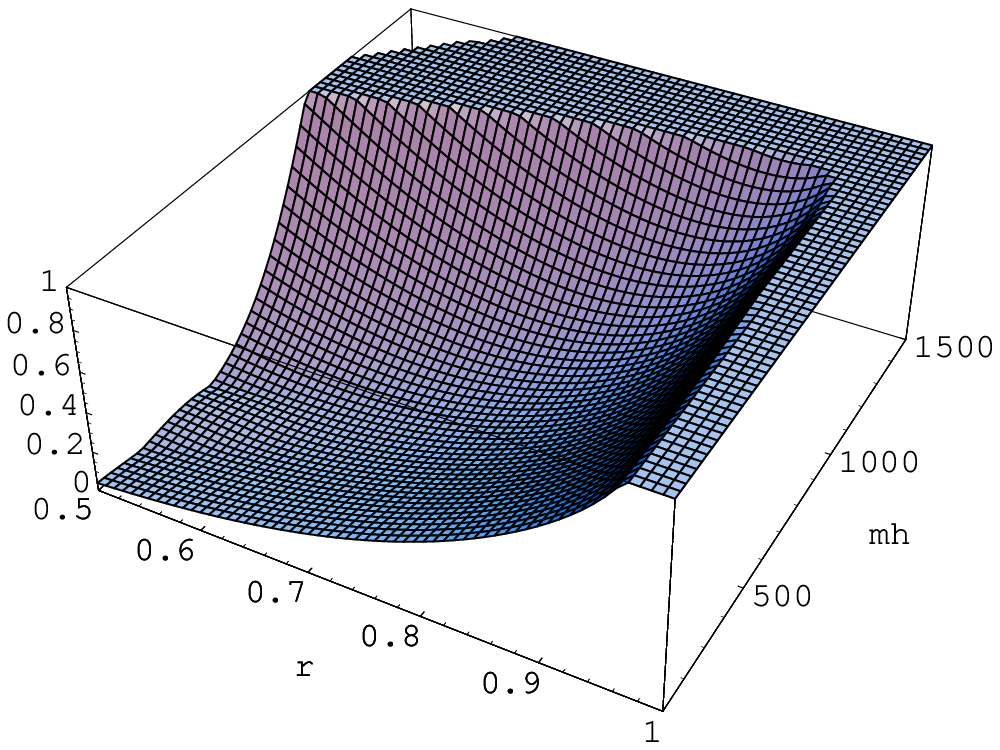 width 7 cm) }
\smallskip
\caption {$|a_0|$ (vertical axis) 
as a function of $r=\sqrt{s}/M_S$ 
and $m_H$ (GeV) for $M_S = 5$ TeV with $n=2$ and 
$n=7$, respectively. Cases a) and b) are shown at left and right, respectively.
The allowed parameter space
 with $|a_0|<1$ are located at the left-lower
corners indicated by the dented regions.}
\end{figure}

\begin{figure}[htb]
\centerline{ \DESepsf(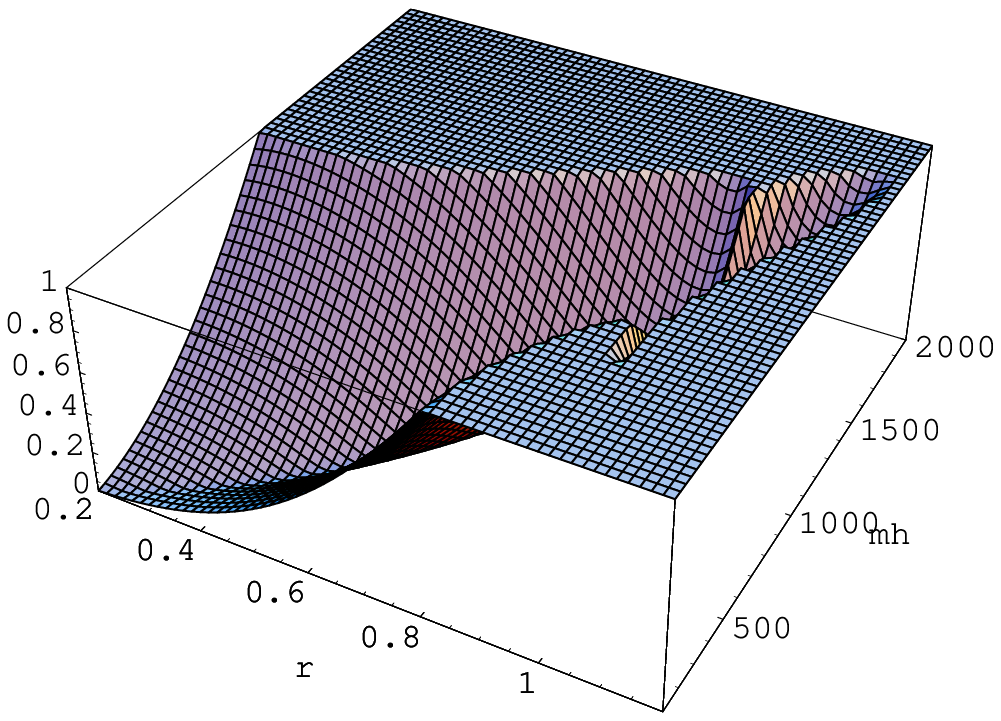 width 7 cm) \DESepsf(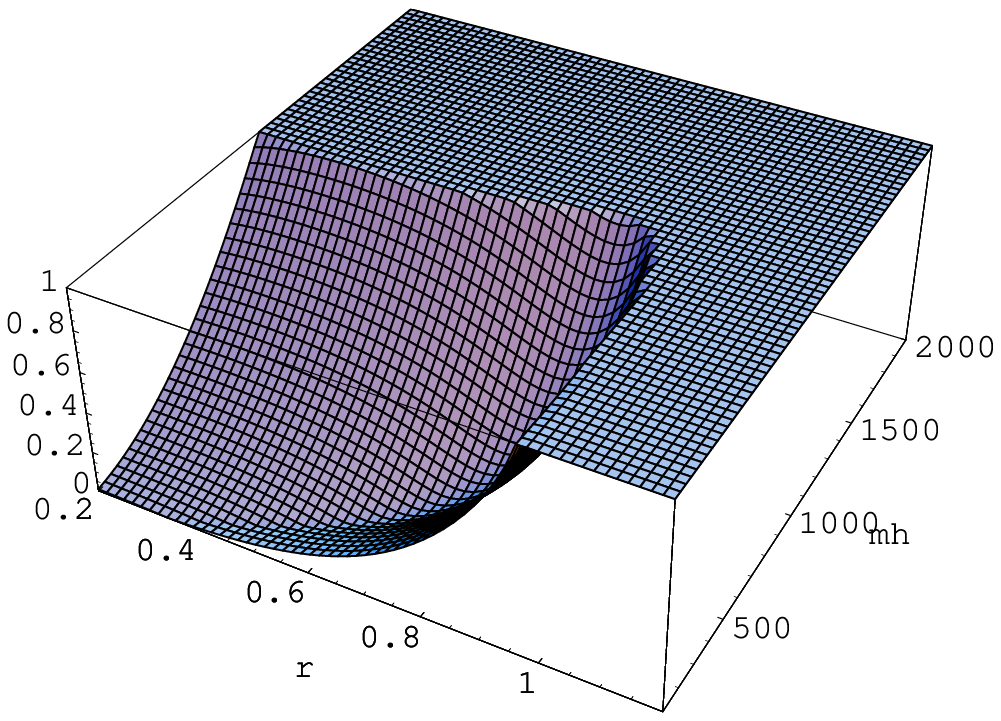 width 7 cm) }
\smallskip
\caption {
The same as Fig. 1, but with $M_S = 100$ TeV. Cases c) and d) are
shown at left and right, respectively.
For $\sqrt{s}$ much smaller than 
$M_S$ ($r \ll 1$) the theory reduces to the SM.}
\end{figure}

\end{document}